\documentclass[3p, authoryear, preprint,review,12pt]{elsarticle}
\usepackage{amssymb}
\usepackage{multirow}
\date{\today}

\begin{document}

\begin{frontmatter}

\title{Taylor's Law and the Spatial Distribution of Urban Facilities}

\author{Liang Wu, Xuezhen Chen, Chunyan Zhao}
\address{School of Economics, Sichuan University, Wangjiang Road 29, 610064 Chengdu, P.R. China}



\begin{abstract}
Taylor's law is the footprint of ecosystems, which admits a power
function relationship $S^{2}=am^{b}$ between the variance $S^{2}$
and mean number $m$ of organisms in an area. We examine the distribution
of spatial coordinate data of seven urban facilities (beauty salons,
banks, stadiums, schools, pharmacy, convenient stores and restaurants)
in 37 major cities in China, and find that Taylor's law is validated
among all 7 considered facilities, in the fashion that either all
cities are combined together or each city is considered separately.
Moreover, we find that the exponent $b$ falls between 1 and 2, which
reveals that the distribution of urban facilities resembles that of
the organisms in ecosystems. Furthermore, through decomposing the
inverse of exponent $b$, we examine two different factors affecting\emph{
}the numbers of facilities in an area of a city respectively, which
are the city-specific factor and the facility-specific factor. The
city-specific factor reflects the overall density of all the facilities
in a city, while the facility-specific factor indicates the overall
aggregation level of each type of facility in all the cities. For
example, Beijing ranks the first in the overall density, while restaurant
tops the overall aggregation level. 
\end{abstract}

\begin{keyword}
Urban Facilities, Spatial Aggregation, Taylor's Law, Factor decomposition.
\end{keyword}


\end{frontmatter}

$\hspace{1.3em}$



$\hspace{1.3em}$



$\pagebreak$

\section{Introduction}

\cite{taylor1961aggregation} describes a power function relationship
$S^{2}=am^{b}$ between the between-sample variance in density $S^{2}$
and the overall mean density $m$ of a sample of organisms in a area,
and shows that the exponent $b$ is specie-specific and concentrates
in the interval $[1,2]$. Moreover, he points out that exponent $b$
can be treated as a clumping index: a) when $b\rightarrow0$, it is
random distribution; a) when $b=1$, it is a Poisson distribution;
and b) when $b$ is significantly larger than $1$, it indicates the
clumping of organisms. Moreover, the concentration of the values of
$b$ in the interval $[1,2]$, indicates that vast majority of the
species follow a clumping and uneven distribution. Taylor's law attracts
wide attention from researchers in various fields and triggers long-lasting
intense discussions, especially regarding the origins and implications 
of the power law and exponent $b$. 

Besides the distribution of organisms, Taylor's power law has found
application to seemingly unrelated phenomena like human sexual pairing
(\cite{andersen1988epidemiological}), human hematogenous cancer metastases, the
clustering of childhood leukemia (\cite{philippe1999scale}), measles epidemics
(\cite{rhodespower}) and gene structures (\cite{kendal2004scale}) etc.,
Given such broad applicability of Taylor's law in many seemly mysterious
and complicated natural processes, one might ask whether there is
any general principle at the basis of all these processes. A large
body of literature has been devoted to this question, and many theoretical
models have been introduced to explain Taylor's Law. For instance,
\cite{anderson1982variability} proposes a Markovian population model and justifies
this model through simulations, which shows that with the increase
in the average population density, the variance to mean relationship
would approach a power function with a maximum exponent value $2$.

Although the research in this field has not yet brought out a unified
theoretical description, it is undeniable that Taylor's law is closely
related to the way of specie proliferation, reflecting the mechanism
for the interaction between the organisms of a specie and the interaction
between the a specie and the ecosystem. %
\footnote{For more detailed and in-depth review of relevant literature about
the application of Taylor's law and the explanation offered, please
refer to (\cite{kendal2004scale}) %
} The concentration of values of $b$ between 1 and 2 implies that
Taylor's law may reflect the joint consequence of certain nonrandom
dynamic processes and random fluctuations of events \cite{linnerud2013interspecific}. In this paper, we are going to explore the potential
relationship between Taylor's law and the distribution of urban facilities. 

Serving as areas for the concentration of human activities, cities
are considered to be the principal engines of innovation and economic
growth (\cite{mumfordcity,hall1998cities}). Today, more than half of the
world population live in cities. The developed world is now
about 80\% urbanization and the entire planet will follow this pattern by
around 2050, with some two billion people moving to cities, especially
in China, India, Africa and Southeast Asia.%
\footnote{UN-Habitat. State of the World’s Cities 2010/2011 — Cities for All:
Bridging the Urban Divide (2010); available at http://www.unhabitat.org%
} Countries around the world are experiencing a rapid urbanization
process, which presents an urgent challenge for developing predictive,
subtle and quantitative theories and methods, providing necessary
technical support for urban organization and sustainable development. 

As consumers of energy and resources and producers of artifacts, information,
and waste, cities have often been compared with biological entities
and ecosystems (\cite{bettencourt2007growth,miller1978living,girardet1996gaia,botkin1997cities}. 
\cite{bettencourt2007growth} shows that there
are very general and nontrivial quantitative regularities of social
activities common to all cities across urban systems, and many diverse
properties of cities, such as patent production, personal income and
crime, are shown to be power law functions of population size. Through
exploring possible consequences of the scaling relations by deriving
growth equations, they quantify the dramatic difference between growth
fueled by innovation versus that driven by economies of scale, suggesting
that as population grows, major innovation cycles must be generated
at a continually accelerating rate, so as to sustain growth and avoid
stagnation or collapse. \cite{bettencourt2010unified} states that cities should
be treated as a complex dynamic system, which is capable of aggregating
and manifesting human cognitive ability, leading to open socioeconomic
development.

In this paper, we obtain the data set of spatial coordinates of 7
facilities in 37 major cities in China, through Baidu Map API. Based
on the data set of accurate location of these facilities, we are able
to explore the micro-structure of these cities and study the distribution
characteristics of urban facilities. We find that there is a power
law relationship $S^{2}=am^{b}$ between the variance $S^{2}$
and mean number $m$ of facilities in the quadrats, and almost all
the values of exponent $b$ fall between 1 and 2. This shows that
similar to the distribution of various species in ecosystems, the distribution
of urban facilities complies with Taylor's law as well. This implies
that the exponent $b$ may reflect the complex mechanism underlying the distribution
of urban facilities and the micro-structure of cities. 
The same facilities in a city may help each other survive, while at the same time, they
compete for various resources, which resembles the relationship between
the organisms of a specie in an area.

Furthermore, in order to study the key factors contributing to the
difference between the values of exponent $b$ and explore the mechanism
underlying the distribution of urban facilities, we decompose the
inverse of exponent $b$ to examine two different factors contributing
to the numbers of facilities in a city respectively. These two factors
are the City-Specific Factor (CSF) and the Facility-Specific Factor
(FSF) respectively. The CSF reflects the overall density of all the
facilities in a city, while the FSF indicates the overall aggregation
level of each type of facility in all the cities. For example, Beijing
ranks first in the overall density, while restaurant tops the overall
concentration level. These findings are consistent with our intuitive
understandings of these cities and urban facilities.

The contribution of this paper mainly lie in the following two aspects.
This is the first paper revealing the relation between Taylor's law
and the distribution of urban facilities. Moreover, we decompose the
inverse of exponent $b$ to examine two different factors contributing
to the numbers of facilities in the quadrats within a city respectively,
and find that both the CSF and the FSF have their own concrete and
specific implications. Furthermore, we discuss some potential factors
underlying the distribution of urban facilities. 

This paper proceeds as follow. Section 2 introduces the source of
the data used in this paper. Section 3 explains our research methods.
Section 4 shows the analytical results. Section 5 decomposes the factors
affecting the number of a facility in the quadrats within a city.
Section 5 discusses the results. Section 7 concludes.

\section{Data Source}
\begin{figure}
\centering
\caption{The distribution of 7 facilities in Beijing.}
\includegraphics[width=0.7\textwidth]{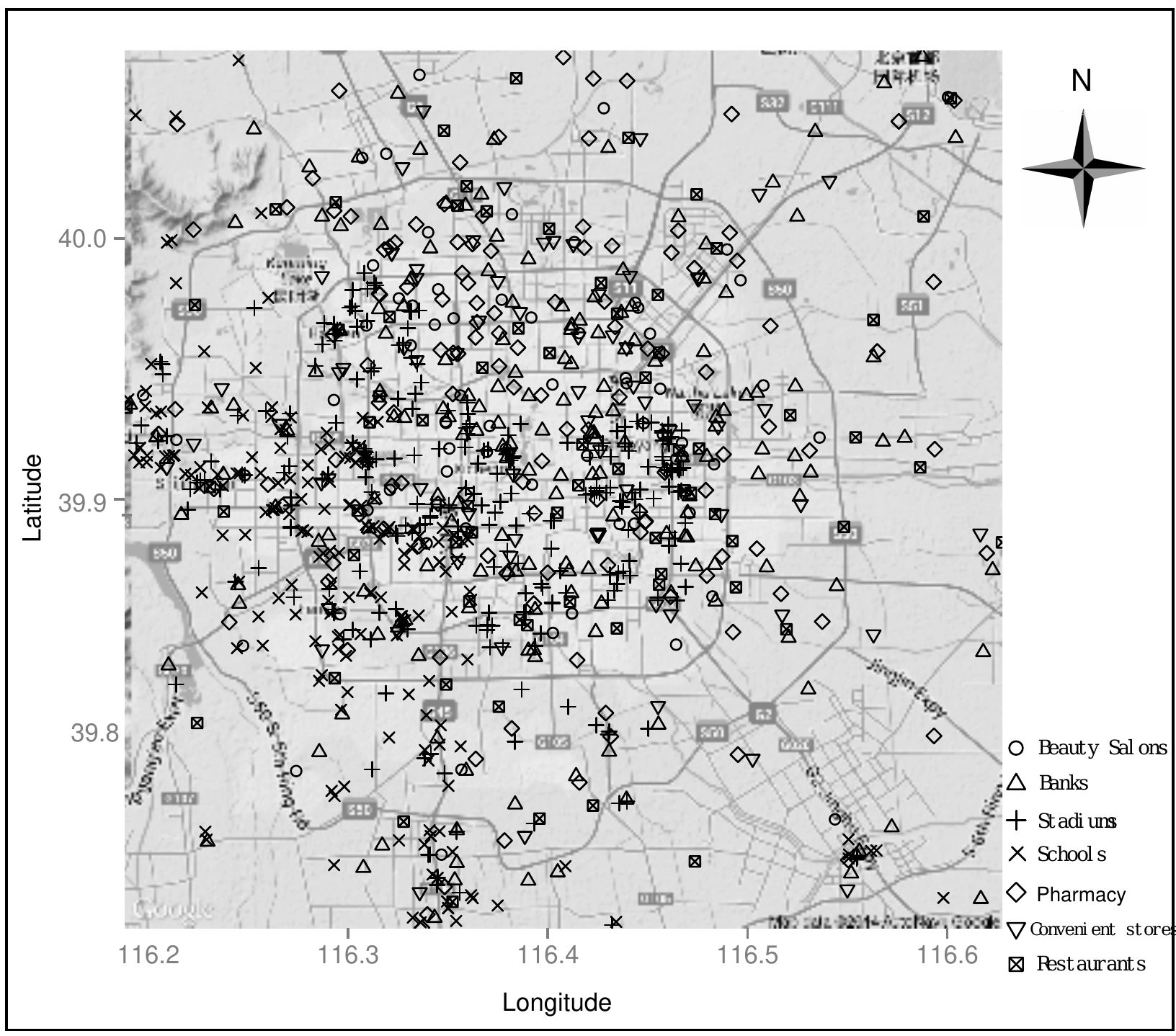}
\label{fig:map}
\end{figure}

$\hspace{1.3em}$Through Baidu Map API, we obtain the data set of
spatial coordinates of 7 facilities, in the city area and adjacent
counties and county-level cities of 37 major cities in China. The
7 facilities are: beauty salons, banks, stadiums, schools, pharmacy,
convenient stores and restaurants. The 37 major cities consist of
4 direct-controlled municipalities (Beijing, Shanghai, Chongqing,
Tianjin), 30 Provincial capitals and sub-provincial cities and 3 other
large cities. The spatial data is the latitude and longitude coordinates
of each facility. For example, as shown in Fig.~\ref{fig:map}, we randomly choose
300 samples for each of the 7 facilities in Beijing and mark these
samples in the map. Then we convert the spatial data of the latitude
and longitude coordinates into the the plane coordinate data denoted
by meter, so as to facilitate the calculation of distance and area
selection. Since it is hard to define the exact boundary of a city,
we fix the central point $(\textnormal{lng}_0,\textnormal{lat}_0)$ of a city,
\footnote{For instance, the latitude and longitude coordinate for the central
point of Beijing is (116.413648E,39.913561N)} 
then calculate the relative distance of each sample to the central
point in the plane coordinate centered at it. For example, given the
latitude and longitude coordinate $(\textnormal{lng}_i,\textnormal{lat}_i)$ of a sample,
its relative distance to the central point can be defined by: $x_i=\textnormal{Distance}((\textnormal{lng}_i,\textnormal{lat}_0),(\textnormal{lng}_0,\textnormal{lat}_0))\times \textnormal{Sign}(\textnormal{lng}_i,-\textnormal{lng}_i)$,
$y_{i}=\textnormal{Distance}((\textnormal{lng}_0,\textnormal{lat}_i),(\textnormal{lng}_0,\textnormal{lat}_0))$ $\times \textnormal{Sign}(\textnormal{lat}_{i},-\textnormal{lat}_{i})$.
Here, $Sing(x)=1$ if $x>0$, 0 if $x=0$, and $-1$ if $x<0$. The
relative distance is equal to the length of arc connecting two points
on a spherical coordinate. The radius of the earth is given by $R=6371004$
meters .

\section{Research Methods}

For each city, we choose a $40km\times40km$ study
area centered at the city central point. This area is then divided
into 16 $10km\times10km$ non-overlapping sub-areas, each of which
is further divided into 25 $2km\times2km$ small quadrats. We denote
the number of a facility in quadrat $i$ of sub-area $j$ with
$X_{ji}$, where $i=1,2,...,25$ and $j=1,2,...16$.
For each of the 16 sub-areas, we calculate the mean and variance of
the number of a facility, thus getting 16 pairs of mean and variance
denoted by $(m_{j},S_{j}^{2})$, where $m_{j}=E[X_{j}]$ and $S_{j}^{2}=Var(X_{j})$
($j=1,2,...16$). In order to get a good estimation
of mean and variance in each sub-area $j$, $X_{i,j}$ should be bigger than 0 for 
a sufficient number of quadrats. However, because of the irregularity
of the city area, a few of the sub-areas could be corresponding with
the depopulated zones, such as sea and mountainous areas. Hence, for
some cities, the numbers of the pairs of means and variances are less
than 16. Nonetheless, this would not affect the robustness of the
analytical results in this paper. 

\begin{figure}
\centering
\caption{G function analysis}
\includegraphics[width=0.7\textwidth]{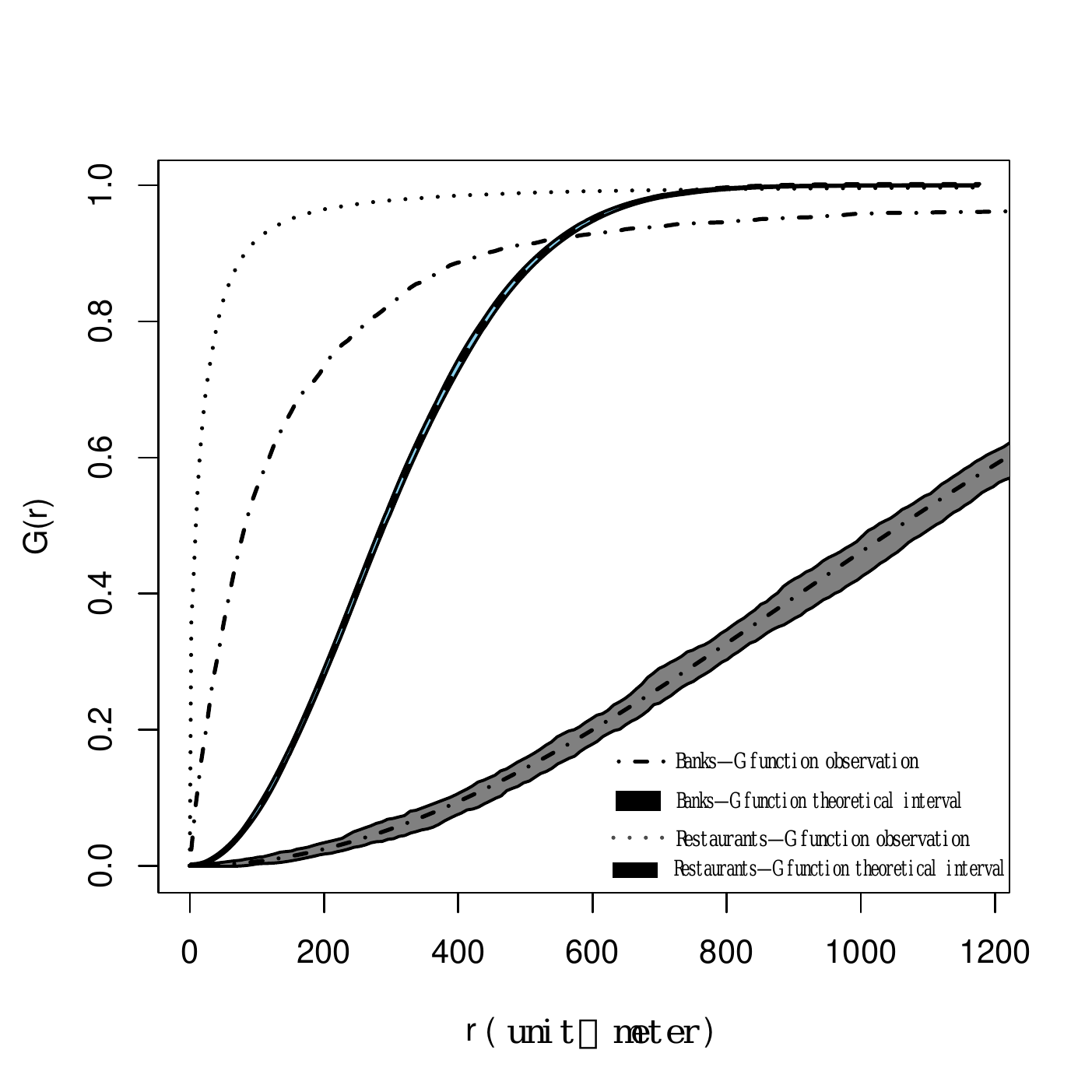}
\label{fig:G}
\end{figure}

\subsection{Descriptive Statistical Analysis of the spatial distribution of the
Facilities}

$\hspace{1.3em}$First of all, we study the statistical characteristics
of the distribution of $X_{ji}$. The main objective is to find out
whether or not they are from the Poisson distribution, based on $\chi^{2}$
test, $G$ function analysis, and Monte Carlo simulation.

\subsection{Taylor's Power Law}
\begin{figure}
\centering
\caption{Aggregated scatterplots for Beauty Salons, Banks, Stadiums and Schools.}
\includegraphics[width=0.7\textwidth]{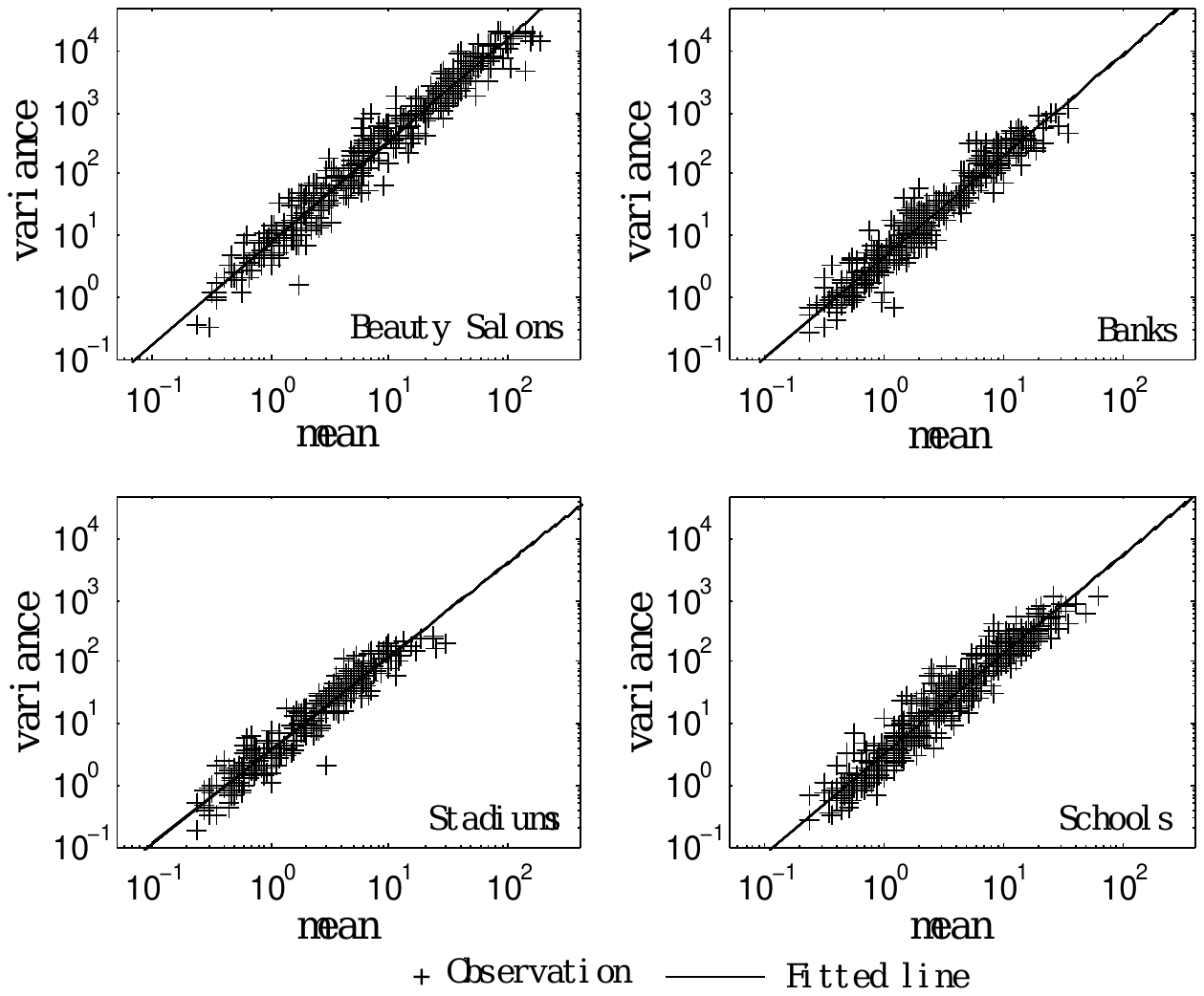}
\label{fig:national}
\end{figure}

\begin{figure}
\centering
\caption{Scatterplots for Beauty Salons, Banks, Stadiums and Schools in Beijing.}
\includegraphics[width=0.7\textwidth]{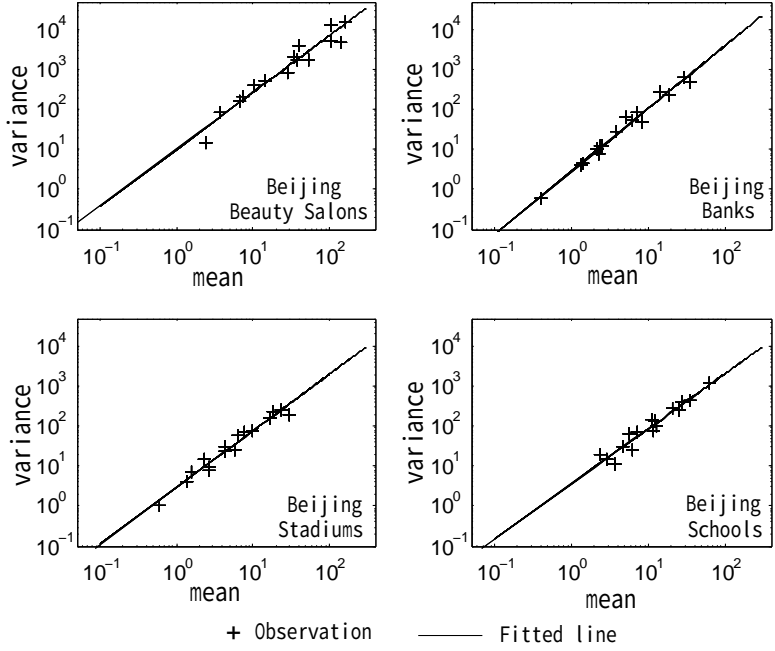}
\label{fig:beijing}
\end{figure}

$\hspace{1.3em}$In order to study the statistical characteristics
of the spatial distribution of the facilities. we will focus on studying
the exponent $b$ in Taylor's power function $S^{2}=am^{b}$. In some
of the cities, the numbers of certain facilities are not large enough,
which may greatly increase the size of error in the estimation of
the means and variances. To solve this problem, for the study of a
facility, we only use the date from the cities which rank in the top
30 in term of the number of this facility. Note that every
pair of means and variances is from an area with the same size to
the other ones'. Hence, the potential relation between the means and
variances only reflects the feature of the events in the areas with
a fixed size, not a scaling rule where the number of events is measured
in a series of expanding areas discussed by \cite{wu2014scaling}.

Through examining whether or not there is linear correlation between
the natural Logarithm values of the means and variances, i.e. $\log S^{2}=loga+b\log m$,
we can judge whether or not Taylor's power law is applicable in the
distribution of the urban facilities.

\begin{table}
\label{tab:national}
\caption{Aggregated parameter fitting results of Taylor’s law.}
\centering
\begin{tabular}{c|c|c|c|c|c|c|c}
\hline
\begin{tabular}{@{}c@{}}Urban \\ Facilities\end{tabular}  & \begin{tabular}{@{}c@{}}Beauty \\ Salons\end{tabular}   & Banks & Stadium & Schools & Pharmacy & \begin{tabular}{@{}c@{}}Convenient \\ Stores\end{tabular}  & Restaurants \\
\hline
$b$ & 1.66 & 1.63 & 1.52 & 1.61 & 1.50 & 1.64 & 1.74 \\
\hline
$\log a$ & 2.03 & 1.52 & 1.30 & 1.16 & 1.55 & 1.55 & 1.82 \\
\hline
\end{tabular}
\end{table}

\section{Analytical Results}

\subsection{Aggregation Test }

By dividing the $40km\times40km$ study area of each city into $2km\times2km$ quadrats, and counting the number of events in each quadrat, we can first apply statistic test to see whether urban facilities are aggregated or not.
First, the $\chi^2$ test is applied against the null hypothesis that urban facilities follow Poisson distribution. For all the 7 facilities in each of the 37 cities, based on the $\chi^{2}$ score of $T_{cc}$, we find that $T_{cc}$ is significantly larger than
the result from the Poisson distribution. Under the significance level
$\alpha<0.001$, we can reject the null hypothesis of Poisson distribution.
For the detailed results for five cities: Beijing, Shanghai, Guangzhou,
Chengdu and Wuhan, please refer to Table A1 in Appendix 1.

We also carry out G function analysis on the spatial distribution
of the facilities in each city. $G$ function is defined as the distribution function of the nearest neighbour distance.
For example, the dashed and the dotted
line in Fig.~\ref{fig:G} represent the G function of the restaurants and that
of the banks in Beijing respectively. 
In order to illustrate that the two considered facilities are aggregated, we plot the empirical G functions together with the theoretic G functions if the facilties are Poisson distributed. 
Suppose that the restaurants and the banks in Beijing follow the Poisson distribution, whose parameters are estimated from the real data by assuming that it is Poisson distributed. Then these estimated parameters can be used to generate 99 groups of sample data for each facility by Monte Carlo simulation.
After calculating the G functions of those sample data, we can get the envelope of those G functions by taking the maximum and minimum at each distance $r$. The envelope of G functions approximate the 99\% confidence interval. The theoretical envelopes of restaurants and banks in Beijing are plotted as the shaded area in Fig.2. We find that the empirical G functions always lie above the confidence interval of the random spatial distribution from
the Monte Carlo simulation. Both dotted line and dashed line in Fig. 2 which represents the restaurants and banks in Beijing rises sharply within the range of 200 meters, which implies that the percentage of other facilities locating close to a given facility is far bigger than that can be predicted by random spatial distribution. The restaurants and banks are aggregated in space.

\begin{figure}
\centering
\caption{The values of exponent b for 21 cities.}
\includegraphics[width=0.7\textwidth]{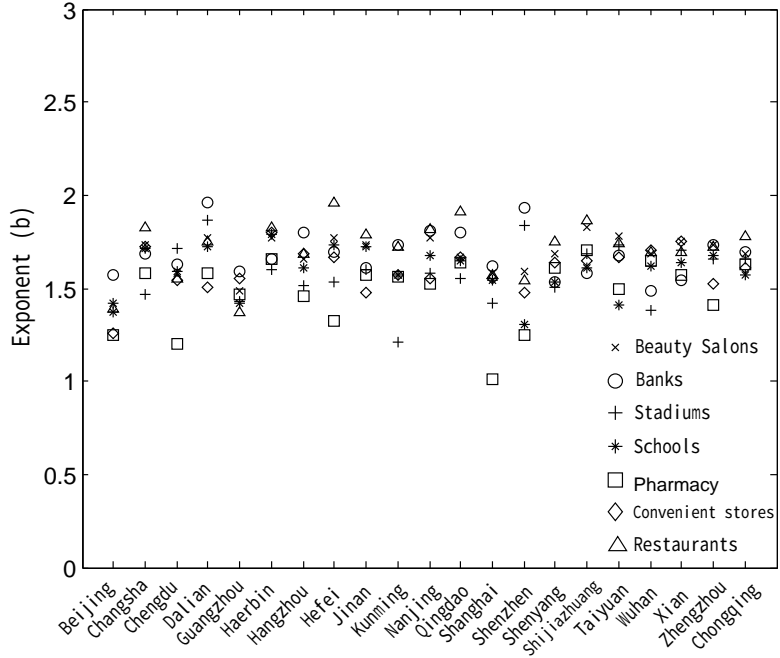}
\label{fig:b}
\end{figure}

\subsection{Taylor's Power Law}

In this subsection, we will examine the aggregated
distribution characteristic of each facility in all 30 cities chosen
for this facility. Figure~\ref{fig:national} shows the scatterplot of the means and
variances for each of the four types of facilities, beauty salons,
stadiums, schools and banks, in 30 cities corresponding with each
of them. The results of the other three facilities (pharmacy, convenient stores, and restaurants) are similar. 
As we can see from this figure, all the means and variances
of each facility are distributed along a straight line in the log-log plot
coordinates, which means they could be described by the power function,
i.e., they satisfy Taylor's power Law. 

Based on the linear regression of the means on the variances, we can
estimate the values of the parameters in Taylor's power function for
each facility. For example, the regression result for schools is:
$\log S^{2}=1.16+1.61\log m$, thus the corresponding intercept and slope
are given by $\log a=1.16$ and $b=1.61$ respectively. $b>1$ indicates
that schools follows the aggregated distribution, which is consistent
with statistical test results in the above subsection. Since
there is no characteristic scale for the power function, this implies
the aggregation degree of schools is similar at various levels of mean concentration. Moreover,
for the t-test on the intercepts and slops,
the significance levels are all well below 0.001. The null hypothesis that either intercept or slope is zero can be rejected. Besides the four
types of facilities in Figure~\ref{fig:national}, the analytical results of the rest
three kinds show that their spatial distributions are all consistent
with Taylor's law. It is worthy noting that values of the intercepts
concentrate in the interval $[1.50,2]$.

In the above, we have examined the aggregated distribution of the facilities
in 30 cities. Now we examine the distribution of the facilities in
a single city. Taking Beijing as an example, we draw the scatterplots
of the means and variances for beauty salons, stadiums, schools and
banks in Figure~\ref{fig:beijing}. As we can see from Figure~\ref{fig:beijing}, for each facility,
16 pairs of the means and variances are distributed
along a straight line in log-log plot, which clearly indicates that it satisfies Taylor's law. 


Based on the linear regression for each facility in a city, the exponent $b$
falls between 1.0 and 2 (see Table~\ref{tab:app}), which
means that within each city, all the facilities follow aggregation
distribution. Once again, the results are consistent with the conclusion
from aggregation test.

The graphs in Fig.~\ref{fig:national} are derived by the stacking of log-log scatters
for all cities. The means and variances are scattered within a band along
a straight line, which reflects the fluctuation in the values
of exponent $b$ in different cities. The main reasons leading to
the fluctuation may lie in the following two aspects. On one hand,
the cities are different from each other in terms of their environmental and socioeconomic
conditions, which may result in the differences in the slops of the power functions. 
On the other hand, when the number
of a facility is not larger enough, the division of a city into 16
sub-areas may lead to considerable estimation errors of means and variances. 

The 7 groups of 30 cities chosen for each facility are different from
each other, and there are only 21 cities which are present in all these 7 groups. In
order to give a clear picture of the estimation values of exponent
$b$ for each facility, we put together the values
of exponent $b$ for each facility in those 21 common cities as in Figure~\ref{fig:b}. As we can see from
Figure~\ref{fig:b}, almost all the values fall between 1 and 2, and concentrate
around 1.6. There are some values outside the interval $\left[1,2\right]$,
this may result from the the estimation error due to insufficient
sample quantity. Despite the differences between the exponent
values, the distribution of every facility still complies with Taylor's
law. Based on the existing study of Taylor's law in ecology, it is
reasonable to infer that there are a particular or a few mechanisms
underlying the applicability of Taylor's law in the distribution of
urban facilities.

\section{Decomposing the Factors Affecting the Number of A Facility}
\begin{table}
\caption{The City-Specific Factor in 21 Cities}
\centering
\begin{tabular}{c|c|c|c|c|c|c|c}
\hline
City & Beijing & Changsha & Chengdu & Dalian & Guangzhou & Haerbin & Hangzhou \\
\hline
$C_i$ & 0.26 & 0.13 & 0.18 & 0.11 & 0.21 & 0.11 & 0.15 \\
\hline
City & Hefei & Jinan & Kunming & Nanjing & Qingdao & Shanghai & Shenzhen \\
\hline
$C_i$ & 0.14 & 0.14 & 0.18 & 0.13 & 0.12 & 0.23 & 0.19 \\
\hline
City & Shenyang & Shijiazhuang & Taiyuan & Wuhan & Xi'an & Zhengzhou & Chongqing \\
\hline
$C_i$ & 0.15 & 0.12 & 0.14 & 0.16 & 0.13 & 0.15 & 0.13 \\
\hline
\end{tabular}
\end{table}
\begin{table}
\label{tab:fsf}
\caption{The Facility-Specific Factor for 7 Facilities}
\centering
\begin{tabular}{c|c|c|c|c|c|c|c}
\hline
\begin{tabular}{@{}c@{}}Urban \\ Facilities\end{tabular}  & \begin{tabular}{@{}c@{}}Beauty \\ Salons\end{tabular}   & Banks & Stadium & Schools & Pharmacy & \begin{tabular}{@{}c@{}}Convenient \\ Stores\end{tabular}  & Restaurants \\
\hline
$f_j$ & 0.44 & 0.44 & 0.49 & 0.48 & 0.53 & 0.47 & 0.43 \\
\hline
\end{tabular}
\end{table}

As we have mentioned in subsection 4.2, the differences
in socioeconomic conditions of the cities and the distinct features
of various facilities may result in the differences between the values
of exponent $b$ in the power functions. In order to explain the fluctuation
in the values of $b$ for various facilities and in different cities,
we can decompose the factor affecting the number of a facility in
the quadrats of a city into two main factors: city-specific factor
and facility-specific factor. Through the ordinary least square regression,
this decomposition may remove a proportion of estimation
error due to insufficient sample quantity, so that we can illustrate
the relative contributions of CSF and FSF to the number of a facility in the quadrats of a city.

Assume: a) the number of a facility in a quadrat of a city is jointly
determined by the CSF and FSF, which are assumed to be independent
from each other; b) these two factors satisfy the following equation for decomposition:
$X_{ij}=Y_{i}+Z_{j}$, where: a) $X_{ij}$ stands for the quantity
of the facility $j$ ($j=1,2,...,7$), in a quadrat of city $i$ ($i=1,2,...,21$);
b) $Y_{i}$ represents the CSF; and c) $Z_{j}$ represents the FSF.
It is clear that the mean value of $X_{ij}$ could be expressed as: 

\begin{eqnarray}
m_{ij}=\textnormal{E}(X_{ij})=\textnormal{E}(Y_{i})+\textnormal{E}(Z_{j})=m_{Y_{i}}+m_{Z_{j}},
\end{eqnarray}
where $m_{Y_{i}}=\textnormal{E}(Y_{i})$ and $m_{Z_{j}}=\textnormal{E}(Z_{j})$, which represent
the average contribution value of the CSF and that of the FSF respectively.
We further assume that there is a power function relation with exponent
$c_{i}$ between the variance $S^{2}$ and the average contribution
value of the CSF, while the power function for FSF
is with exponent $f_{i}$. This assumption can be represented by the
following two equations:

\begin{eqnarray}
m_{Y_{i}}=(S_{ij}^{2})^{c_{i}}/\tilde{a}, \\
m_{Z_{i}}=(S_{ij}^{2})^{f_{i}}/\tilde{a}
\end{eqnarray}

Here, it is assumed that there is a common factor $\tilde{a}$
in the above two power functions. It should be noted that when $S_{ij}^{2}$(or
$m_{ij}$) is sufficiently large, these two power functions will play a dominant role, making this assumption relatively trivial.
By combining equations (2) and (3), we have:

\begin{eqnarray}
\frac{m_{Y_{i}}}{m_{Z_{i}}}=S_{ij}^{2(c_{i}-f_{i})},
\end{eqnarray}

which means that the relative average contribution, between the CSF
and the FSF, is determined by the difference between the exponents
$c_{i}$ and $f_{i}$. Particularly, when $S_{ij}^{2}\gg1$ (or $m_{ij}\gg1$)
is sufficiently large, the factor corresponding to the relative larger
exponent between $c_{i}$ and $f_{i}$, will play an absolutely dominant
role in the average contribution to the number of a facility in an
area. For instance, when $f_{i}>c_{i}$, the FSF plays an absolutely
dominant role.Based on the Taylor's power function $S^{2}=am^{b},$
we can derive:
\begin{eqnarray}
m_{ij}=(S_{ij}^{2})^{\frac{1}{b_{ij}}}/a^{\frac{1}{b_{ij}}}
\end{eqnarray}

Hence, based on Eqs. (1), (2), (3) and (5), we can infer:
\begin{eqnarray}
(S^{2})^{\frac{1}{b_{ij}}}=(S_{ij}^{2})^{c_{i}+f_{i}},
\end{eqnarray}

thus,
\begin{eqnarray}
\label{Eq:deO}
\frac{1}{b_{ij}}=c_{i}+f_{i}
\end{eqnarray}

Considering the errors in the data, we can rewrite Eq.~(\ref{Eq:deO}) as:
\begin{eqnarray}
\label{Eq:deE}
\frac{1}{b_{ij}}=c_{i}+f_{i}+\varepsilon_{ij}.
\end{eqnarray}

Based on Eq.~(\ref{Eq:deE}), we can define the objective function:
\begin{eqnarray}
J=\sum_{i=1}^{n}\sum_{j=1}^{m}\left(\frac{1}{b_{ij}}-(c_{i}+f_{i})\right)^{2},
\end{eqnarray}
where $n$ and $m$ represent the number of cities and that of a facility
respectively. Through minimizing the above objective function, we
can derive the value of $c_{i}$ and that of $f_{i}$. We find that
$E[|\frac{1}{b_{ij}}-(c_{i}+f_{i})|/\frac{1}{b_{ij}}]=2.7\%$, which
means that $c_{i}$ and $f_{i}$ jointly contribute $97.3\%$ to the
value of $\frac{1}{b_{ij}}$. The vast majority of the values of $c_{i}$
fall between 0.1 and 0.26, contributing around 25\% to the value of
$\frac{1}{b_{ij}}$; while the values of $f_{i}$ are rather concentrated
in a small interval $[0.43,0.53]$, contributing about $75\%$ to
the value of $\frac{1}{b_{ij}}$. Hence, it is obvious it is the FSF
that plays a dominant role in determining the number of a facility
in an area within a city. Based on the above analysis, we can infer
that when the number of a facility in an area is relative large, it
is largely the FSF that leads to the agglomeration of the facility;
while the role of city-specific factor is minor. 

In order to better understand the above analytical results, let's
look at the case of beauty salons in Beijing. The relevant values
for this case is given by: $c=0.26$, $f=0.44$, $\log(a)=2.23$ and
$b=1.42$. When the variance $S^{2}=5000$, the mean value of this
facility in a quadrat in Beijing is: $m=(S^{2})^{\frac{1}{b}}/a^{\frac{1}{b}}$.
The average contribution of the CSF for Beijing is: $m_{Y}=(S^{2})^{c}/a^{\frac{1}{2b}}$,
and the average contribution of the FSF is: $m_{Z}=(S^{2})^{f}/a^{\frac{1}{2b}}$.
Based on these values, it is clear that the FSF plays a dominant role
in the mean value of beauty salons in a quadrat in Beijing. In the
following two tables, we list the values of the contribution factors
for all the 7 facilities in 21 cities. 

As we can see from Table 2, the values of the CSF vary significantly
among different cities. These values directly determine the mean
value for all facilities in a quadrat within a city, hence they can
be seen as the indicators of the overall density of all the facilities
in their corresponding cities. The CSF for Beijing is 0.26, which
shows the highest density of the facilities among all the cities.
It is followed by Shanghai (0.23), Guangzhou (0.21). Dalian and Haerbin
have the same lowest value of the CSF at 0.11. 

For the values of the
FSF, we need to understand from another perspective. Because $\frac{1}{b_{ij}}=c_{i}+f_{j}$,
for given a value of $c_{i}$ in a city, the smaller the value of $f_{j}$, the
larger the values of $b_{ij}$ and therefore larger $S_{ij}^{2}$ for a given mean $m_{ij}$ in the corresponding
city. Larger variances imply greater differences between the numbers
of facility $j$ in different quadrats. At some place, the number is small, but at another place, the
number can be very large which means that the facility tend to aggregate in space. 
On the contrary, when $f_{j}$
becomes larger, given the same value of mean, the variance falls,
thus the distribution tends to behave more like the Poisson distributio, which implies
a weaker aggregation.
In our decomposition, the value of the FSF for restaurants is 0.43, which is the smallest,
and that for pharmacy is 0.53, which is the largest. This shows
restaurants have the highest degree of aggregation; while pharmacy
has the lowest degree of aggregation.
Restaurants with different styles can coexist at one place, however, due to its interchangeability, 
pharmacy tends to distribute away from each other.

\section{Discussion}

Industries are often geographically concentrated
in particular cities or metropolitan areas, and there are many theories
explaining why the concentration may occur (\cite{marshall2006industry,krugman1990increasing,ellison1999geographic}. 
According to \cite{marshall2006industry}, industries
agglomerate mainly because of the following factors: (i) benefiting
from mass-production (internal economies or scale economies), ii)
saving transport costs by proximity to input suppliers or final consumers,
iii) allowing for labor market pooling, iv) facilitating knowledge
spillovers by reducing communication cost, and v) capitalizing from
the existence of modern infrastructures. \cite{ellison1999geographic}
assess the importance of natural advantage to geographic concentration,
and find that one-quarter of industrial concentration can be explained
by observable sources of natural advantage. \cite{krugman1990increasing} develops
a model of labor market pooling, illustrating how labor market pooling
leads to industrial agglomeration. \cite{audretsch1998agglomeration} states that ‘knowledge
is generated and transmitted more efficiently via local proximity,
economic activity based on new knowledge has a high propensity to
cluster within a geographic region’. 

The analytical results in this paper are in line with the findings
in the literature mentioned in the above paragraph. Beijing, Shanghai
and Guangzhou are generally acknowledged as the three most urbanized
cities in China, and our analytical results show that these three
cities have the highest level of concentration of urban facilities.
\cite{glaeser2004cities} show evidence that services tend to be
located in dense areas because they are more dependent on proximity
to costumers than manufacturing industries. \cite{kolko2007agglomeration} states that
services are less agglomerated but more urbanized than manufacturing.
Moreover, there is a strong tendency of service industries to locate
near their suppliers and customers, because the costs of delivering
services are much higher than the costs of delivering goods. City
streets enable service providers to readily link with large numbers
of their diverse customers, hence they are a good setting for services.
\cite{waldfogel2008median} reveals that there is a strong pattern of retail
establishment sectors, such as restaurants and media, to locate near
demographic groups that regularly buy from that sector. 

\subsection{Animal Grouping Behavior and Facility Aggregating Pattern}

As we have shown in the above analyses, the distribution of urban
facilities resembles that of the organisms in ecosystems. Organisms
feed on various resources, while facilities ‘feed on’ consumer demand.
Organisms are prone to form groups, but the size of group varies between
different species. For example, zebras and wildebeest form large herds,
while the lions usually live in small groups. Urban facilities tend
to agglomerate in an area, while as we can see from Table 3, the degree
of agglomeration varies between different facilities. For instance,
the value of the FSF for restaurants is 0.43, which is the smallest,
and that for pharmacies is 0.53, which is the largest. This shows
restaurants have the highest degree of agglomeration; while pharmacies
have the lowest degree of agglomeration (or highest degree of dispersion).
The same facilities in an area may help each other survive, but at
the same time, they compete with each other in various aspects(such
as consumer demand and raw materials), which resembles the relationship
between the organisms of a specie in an area. 

Considering the similarities between organisms and urban facilities,
we may get a useful guideline in the study of the reasons driving
the agglomeration of facilities, through looking at the factors contributing
to the concentration of organisms.

i) Some organisms agglomerate in an area, because their food is clumped.
As long as there is no pending threat, animals will stop moving and
searching when they reach an area with abundant food. Animals of the
same species survive on the same food, thus an area abundant with
such food will attract them to move from other areas. As a result,
they start to concentrate in this area. Examples include various herbivores,
such as zebras, chinkaras and fallow deers. 

Across services, \cite{kolko2007agglomeration} finds a positive relationship between
urbanization and concentration, and the services that are most likely
to benefit from geographic connections to diverse urban populations,
are also most likely to concentrate. Many urban facilities, such as
restaurants, beauty salons, stadiums and schools, deliver face-to-face
services or serve as venues of meeting people, while the costs of
moving people across space is much higher than that of delivering
products. Hence, urban facilities would concentrate in those areas
where their customers are concentrated. For instance, the restaurants
tend to agglomerate in those business areas with large numbers of
visitors, i.e. their customers are ‘clumped’.

ii) The larger a group, the less likely it is an individual will be
eaten, because your risk is diluted by those around you. Grouping
has been widely accepted as a mechanism for protection from predation.
Moreover, \cite{mooring1992animal} point out that grouping applies
as much to protection of animals from flying parasites as protection
from predators, as long as the encounter-dilution effect works. They
state: ‘the encounter-dilution effect provides protection when the
probability of detection of a group does not increase in proportion
to an increase in group size (the encounter effect), provided that
the parasites do not offset the encounter effect by attacking more
members of the group (the dilution effect)’.

If a number of certain facilities have been surviving in an area,
their success signals to the potential enters that this area is a
desirable location for their business, thus it is less likely that
their business would fail. Such a signaling effect may attract more
and more same facilities into this area or adjacent areas, thus urban
facilities agglomerate. Moreover, when the number of adverse incidents,
such as robbery and extortion, does not increase in proportion to
an increase in group size, an individual facility could become less
likely to suffer from those adverse incidents. This ‘dilution effect’
could be another reason driving the agglomeration of urban facilities.

iii) By living in groups, many prey species become faster in detecting
approaching predators and stronger in defending against them. Goshawks
are less successful in attacks on pigeons in large flocks mainly because
the pigeons are much more alerted to a pending attack sufficiently
early to fly away. Fish swim together in schools to warn each other
of approaching danger. Moreover, when many group-living animals are
confronted with predators, they will resort to mobbing behavior to
defense themselves. 

Urban facilities agglomerating in an area are facing similar market
conditions and potential negative shocks, such as unexpected changes
in consumer preference. By observing what is happening to their peers
in the same industry and communicating with them, the owners of the
facilities may be able to detect the changes in the market earlier.
Such a knowledge spillover effect could enable them to adapt to the
changes in time, so as to improve business performance or reduce loss.
Moreover, when there exist certain group defense mechanisms between
the same facilities in an area, each individual facility will become
less vulnerable to potential threats. For example, the restaurant
owners may cooperate with each other to exert greater pressure on
the supplies who are planning to increase the prices of raw materials. 

iv) For many organisms, the success of finding food is directly related
to the size of group. The desert locusts usually move in swarms of
immense size, which enable them to spot forage efficiently. Naked
mole-rats live in groups of up to 300, and they establish co-ordinated
digging teams in their attempts to find hidden food below ground.
\cite{couzin2005effective} finds that the larger the group, the smaller the
proportion of informed individuals needed to guide the group on the
move. Moreover, only a very small proportion of informed individuals
is required to achieve great accuracy of direction. 

For many urban facilities, the success of attracting more customers
is directly related to the size of group.When the number of certain
facility grows in an area, this may work as a ‘free advertisement’
to the population in this area and those adjacent areas. Large numbers
of same facilities in an area usually imply that the consumers are
entitled to more differentiated services and possible lower prices,
especially for those monopolistic competition industries (such as
restaurants and beauty salons). This may attract more customers from
this area and other areas. Moreover, As we have mentioned in the above,
many urban facilities are served as venues for meeting people. When
people are discussing about where to meet, it is more likely that
they would agree on the location where the facilities concentrate,
instead of those areas with only a few such facilities. Examples include
restaurants and fashion shops.

Animals can benefit from the positive side of grouping, while at the
same time, they have to live with the negative side. Animals may face
some threats resulting from grouping. The dawn cacophony at a sociable
weaver nest attracts various predators, including cape cobra. Individuals
within groups also share adverse attributes such as parasites and
diseases. Moreover, it is paradoxically yet naturally that the very
group members that assist you to either find or subdue your food,
will instantly become your greatest rivals when the time comes to
eat the spoils. Naked mole-rats co-ordinated in digging for food below
ground, nonetheless once food is found, the veneer of co-operation
gives way to conflict immediately.

When the number of restaurants grows in an area within a city, this
may attract more customers from other areas; while at the same time,
a restaurant will have to compete with more rivals for the customers
in this area. \cite{belleflamme2000economic} state ``when the firms in the
same industry are agglomerated, they will have to face the prospects
of tough price competition''. This conclusion could also be applied
to those facilities in the service industry. Moreover, 
\cite{belleflamme2000economic}argue that the price competition can be relaxed through
product differentiation. Generally speaking, the products or services
are more differentiated in the restaurants, this could explain why
the restaurants are more agglomerated than other types of facilities.

\subsection{External Economy vs. External Diseconomy of Agglomeration}

\begin{figure}
\centering
\caption{External economy vs. external diseconomy.}
\includegraphics[width=0.7\textwidth]{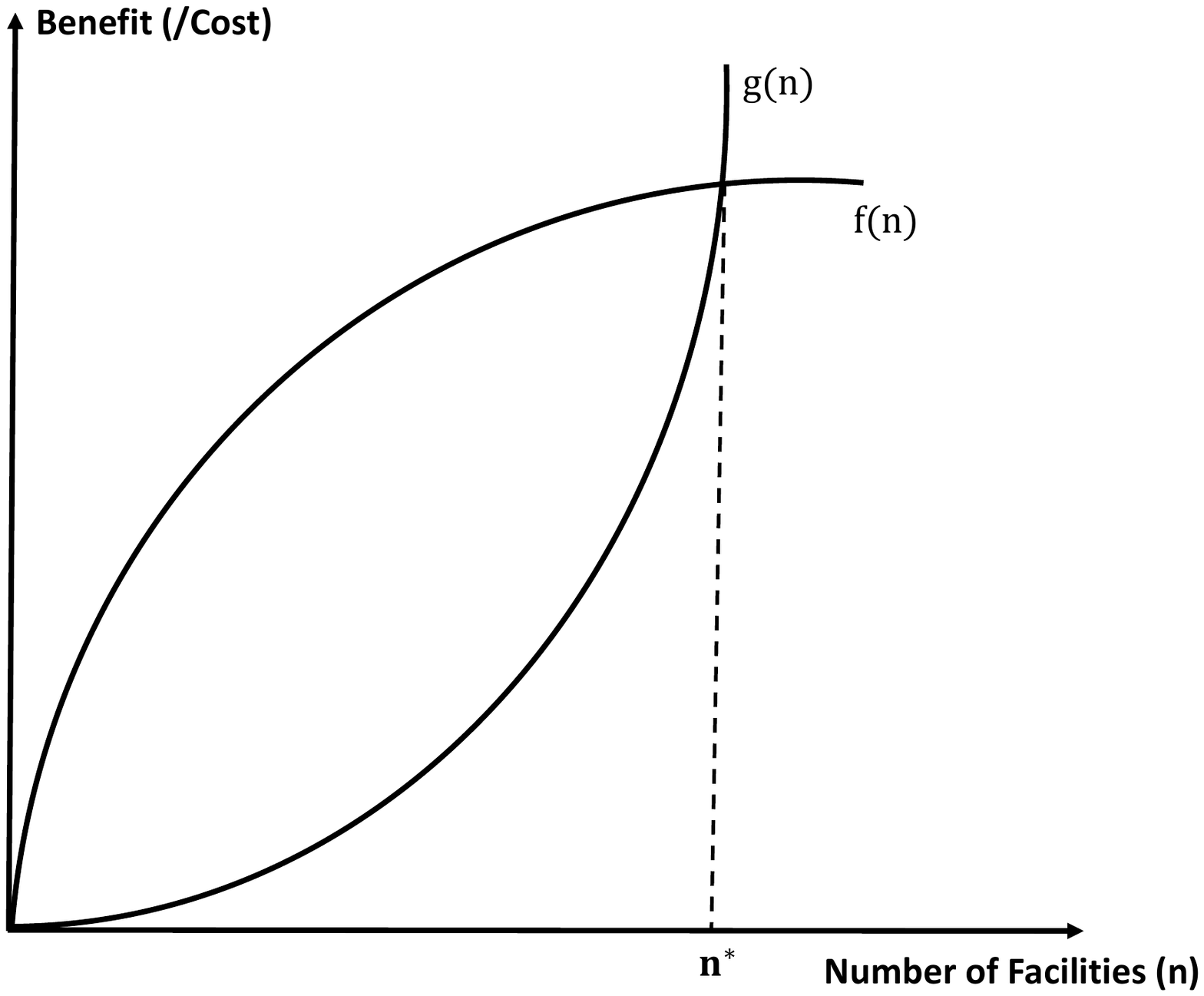}
\label{fig:Exter1}
\end{figure}

\begin{figure}
\centering
\caption{Equilibrium levels for different facilities.}
\includegraphics[width=0.7\textwidth]{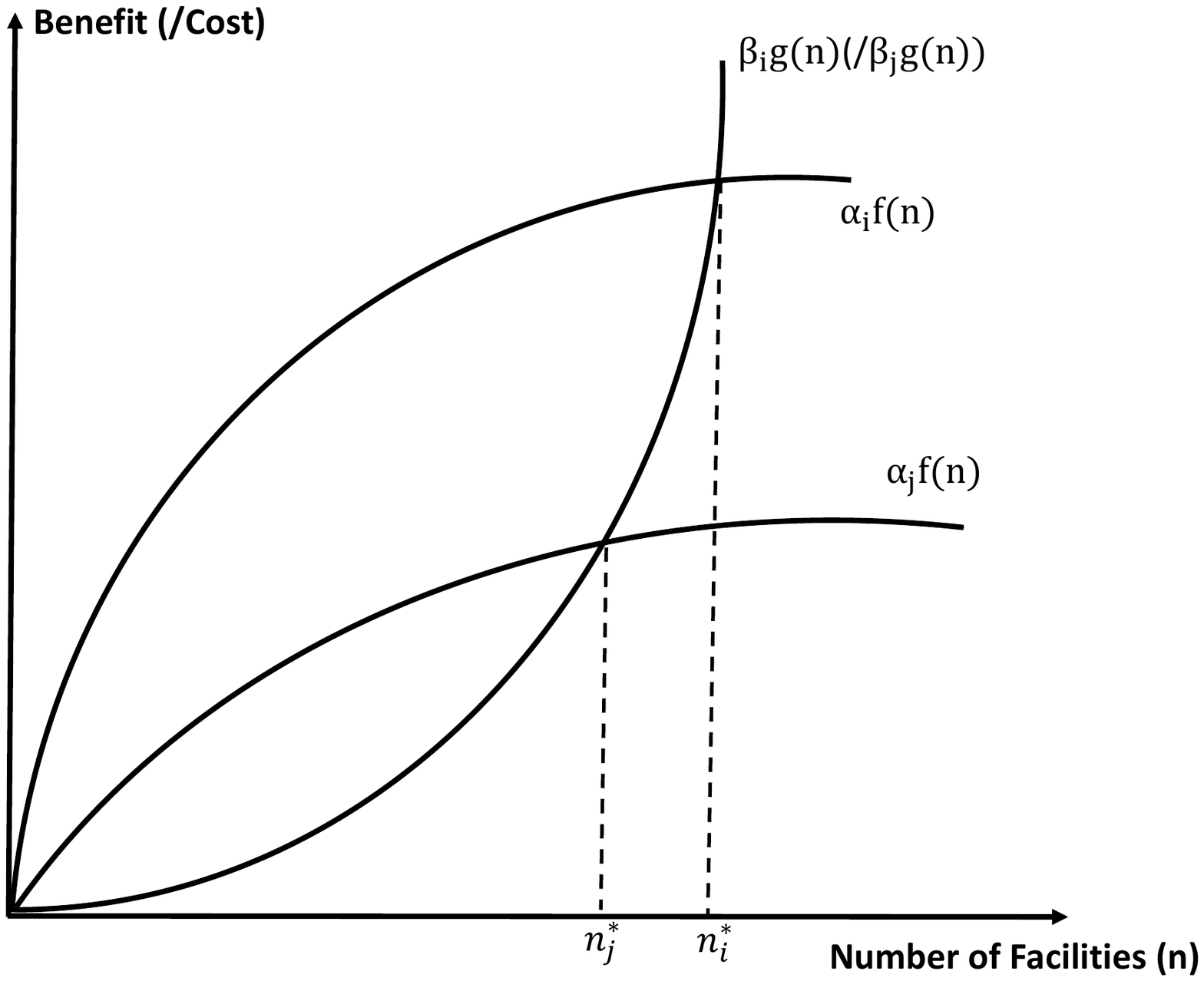}
\label{fig:Exter2}
\end{figure}

\begin{figure}
\centering
\caption{Equilibrium levels of clustering across areas.}
\includegraphics[width=0.7\textwidth]{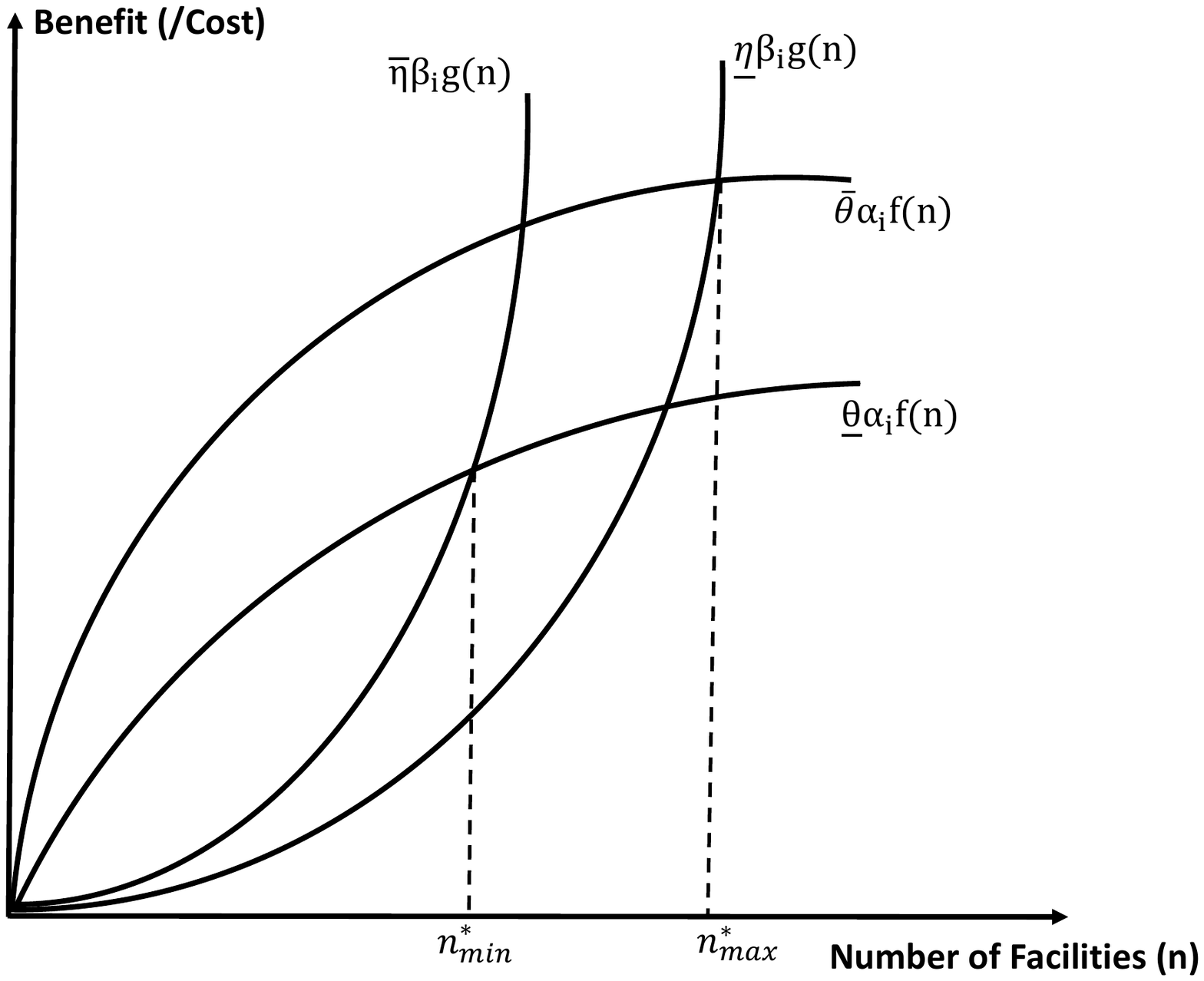}
\label{fig:Exter2}
\end{figure}

The theory of agglomeration economies proposes that firms enjoy positive
externalities from the spatial concentration of economic activities.
These benefits can arise from both intra- and inter-industry clustering
of economic activities (see \cite{fujitavenables,fujita2002economics}).
\cite{fujita2002economics} state: ‘Intuitively the equilibrium spatial
configuration of economic activities can be viewed as the outcome
of a process involving two opposing types of forces, that is, agglomeration
(or centripetal) forces and dispersion (or centrifugal) forces...an
interesting model of economic geography must include both centripetal
and centrifugal forces. The corresponding spatial equilibrium is then
the result of a complicated balance of forces that push and pull consumers
and firms until no one can find a better location’.

In the context of economics, we may use external economy and external diseconomy of clustering to describe the opposite impacts of increasing the number of the same type of facilities in an are. In order to study the distribution pattern of urban facilities, and the differences between the distribution patterns of various urban facilities, we need to explore the factors leading to external economy or external diseconomy, and examine the differences between these factors and how they affect the patterns. Facilities such as restaurants and beauty salons are more concentrated than some other facilities such as pharmacies and stadiums, because for the former, external economy dominates external diseconomy over larger numbers of agglomerated facilities, compared with the later. 

A potential entrant is determining whether or not to locate his business in an area with n incumbent, where $n\geq 0$.   Suppose the external economy will bring extra $f(n)$ units of profit to his business, while the external diseconomy will reduce the profit by $g(n)$ units. Here, let's assume: i) $f(n)$ is concave function with $f(0)=0$, $f′(0)=\infty$, $f′(n)>0$ and $f′'(n)<0$ for all $n>0$; ii) $g(n)$ is a convex function with $g(0)=0$, $g′(0)=0$, $g'(n)<0$ and $g′′(n)>0$, for all $n>0$. We may normalize the profit from locating the business in other areas to zero, then the decision rule of the entrant is reduced to comparing the benefit of external economy and the cost of external diseconomy.

In Figure~\ref{fig:Exter1}, the unique crossing point of $f(n)$ and $g(n)$ is corresponding to the number $n^{*}$. It is clear that when the number of incumbents in an area is $n^{*}$, the potential entrant is indifferent between entering this area or not. When the number of incumbents is smaller than $n^{*}$, the potential entrant will enter. When the number of incumbents is larger than $n^{*}$, the potential entrant will not enter; Hence, given the the above simple setting, the equilibrium level of clustering is $n^{*}$ for the same type facilities in an area. 

The degrees of agglomeration vary across different facilities, which means the equilibrium levels of clustering are different between facilities. This results from different extent of external economy and external diseconomy. Suppose there are l types of facilities. In order to reflect such a difference, we can introduce a parameter $\alpha_{i}$, where $\alpha_{i}>0$  and $i=1,2,...l$, to scale up or scale down external economy; and a parameter $\beta_{i}$, where $\beta_{i}>0$, to scale up or scale down external diseconomy. For type $i$ facilities, the external economy and external diseconomy are represented by $\alpha_{i}f(n)$ and $\beta_{i}g(n)$ respectively. There is an unique crossing point of $\alpha_{i}f(n)$  and $\beta_{i}g(n)$ corresponding to $n_{i}^{*}$, which is the equilibrium level of clustering for type $i$ facilities. By comparing the $n_{i}^{*}$  for different types, we can find out which type has higher degree of agglomeration. For example, if $\beta_{i}=\beta_{j}$, while $\alpha_{i}>\alpha_{j}$, where $i\neq j$, then it is easy to show $n_{i}^{*}>n_{j}^{*}$ (see Figure~\ref{fig:Exter2}).

As we have shown, the concentration level of a facility varies across cities and different areas within a city. Certain specific features of an area, such as population density and location, may have significant impact on the the concentration level of a facility. Suppose there are h
  areas, we may introduce parameter $\theta_{ij}$
  and $\eta_{ij}$
 , where $j=1,2,...,h$
 , to represent the area-specific impact on external economy and external diseconomy for facility $i$
 , where $i=1,2,...l$
 . Here, $\theta_{ij}\in[\underline{\theta_{i}},\overline{\theta_{i}}]$
  and $\eta_{ij}\in[\underline{\eta_{i}},\overline{\eta_{i}}]$
 , where $\overline{\theta_{i}}>\underline{\theta_{i}}>0$
  and $\overline{\eta_{i}}>\underline{\eta_{i}}>0$. Then $\theta_{ij}\alpha_{i}f(n)$
  and $\eta_{ij}\beta_{i}g(n)$
  stand for the external economy and external diseconomy for facility $i$
  in area $j$
  respectively, where $i=1,2,...l$
  and $j=1,2,...,h$
 . It is clear that for facility $i$
  in area $j$
 , the lowest equilibrium level of clustering $n_{\textnormal{min}}$
  is jointly determined by $\underline{\theta_{i}}\alpha_{i}f(n)$
  and $\overline{\eta_{i}}\beta_{i}g(n)$
 , while the highest level $n_{max}$
  is jointly determined by $\overline{\theta_{i}}\alpha_{i}f(n)$
  and $\underline{\eta_{i}}\beta_{i}g(n)$
 . This can be shown in figure 8.

\section{Conclusion }

Based on the data set of spatial coordinates of 7 facilities in 37 major cities in China, we explore the micro-structure of these cities and study the characteristics of the distribution of urban facilities. We find that there is a power law function relationship $S^{2}=am^{b}$ between the variance $S^{2}$ and mean number $m$
  of facilities in the quadrats, and almost all the values of exponent $b$
  falling between 1 and 2. This shows that the distribution of urban facilities complies with Taylor's law. The same facilities in a city may help each other survive, while at the same time, they compete for various resources, which resembles the relationship between the organisms of a specie in an area. 

Furthermore, in order to study the key factors contributing to the difference between the values of exponent $b$ and explore the mechanism underlying the distribution of urban facilities, we decomposing the inverse of exponent $b$
  into two different factors contributing to the numbers of facilities in a city respectively: the CSF and the FSF. we find that the values of the CSF vary significantly between different cities, and different facilities have different degree of agglomeration. It is interesting to note that Beijing, Shanghai and Guangzhou, the three largest and most developed cities in mainland China, show the highest density of the facilities among all the cities. Moreover, restaurants have the highest degree of agglomeration; while pharmacies have the lowest degree of agglomeration. These findings are consistent with our intuitive understandings of these cities and urban facilities.

Spatial Statistics has been becoming more and more important in urban and regional studies, which greatly contributes to the development of theories and knowledge in this field. Through studying the characteristic of the spatial distribution of urban facilities, we can find out whether or not there is an agglomeration property of a facility and how strong it is. \cite{bettencourt2010unified} state: ‘...cities are remarkably robust: success, once achieved, is sustained for several decades or longer, thereby setting a city on a long run of creativity and prosperity’. While the distribution of urban facilities may play a critical role in setting a city on a long run of creativity and prosperity. 

It is important for us to carry out further studies on the distribution of urban facilities, and the potential directions could lie in the following three aspects. Firstly, through combining spatial statistics, economic theories and other relevant fields, we could further explore the rationals and mechanisms underlying the distribution of urban facilities, and examine its impact on socioeconomic development in a city and the adjacent regions. Secondly, when we have sufficient panel data, we could examine the evolution of the distribution of urban facilities over both the time and space, and explore the relationship between the evolution process and the changes in socioeconomic development indicators, such as income per capita, population density and health indicator, etc. Lastly, we could explore relevant theoretical frameworks that could help improve the distribution of urban facilities, thus facilitating sustainable development of cities. 

\section*{Acknowledgements}
The partial financial support from the Fundamental Research Funds for the Central Universities 
under grant number skyb201403, and the Start up Funds from Sichuan University under grant number yj201322 is 
gratefully acknowledged. The other co-author Xuezhen Chen would like to acknowledge the financial support 
from Sichuan University: No. YJ201353 and No. Skyb201404.

\bibliographystyle{elsarticle-harv} 
\bibliography{urban}

\section*{Appendix }

\begin{table}
\label{tab:app}
\caption{The Analitical Results of 7 urban facilities in Five Representative Cities}
\centering
\begin{tabular}{c|c|c|c|c|c|c|c|c}
\hline
City & Ss.
 & \begin{tabular}{@{}c@{}}Convenient \\ Stores\end{tabular} & Restaurants  & \begin{tabular}{@{}c@{}}Beauty \\ Salons \end{tabular} & Stadiums & Schools & Pharmacy & Banks \\
 \hline
 \multirow{4}{*} {Beijing}
 &$n$ & 9142 & 56844 & 25405 & 3768 & 6660 & 3719 &  3521 \\
 &$T_{cc}$ & 11312.2 & 122323.9 & 82491.4 & 6841.6 & 12257.4 & 3968.5 & 9261.9 \\
 &$p$ & (0.000) & (0.000) & (0.000) & (0.000) & (0.000) & (0.000) & (0.000) \\
 &$b$ & 1.25& 1.39& 1.42& 1.42& 1.37& 1.25& 1.57 \\
 \hline 
  \multirow{4}{*} {Shanghai}
  &$n$ & 14038 & 59065 & 23194 & 3079 & 5519 & 2939 & 2923 \\
  &$T_{cc}$ & 26667.0 & 170349.1 & 77446.1 & 6935.5 & 10858.1 & 3534.1 & 10940.6 \\
  &$p$ & (0.000) & (0.000) & (0.000) & (0.000) & (0.000) & (0.000) & (0.000) \\
  &$b$ & 1.64 & 1.75 & 1.78 & 1.51 & 1.61 & 1.50 & 1.58 \\
  \hline 
  \multirow{4}{*} {Guangzhou}
   &$n$ & 7575 & 25824 & 11626 & 1411 & 3784 & 4408 & 2063 \\
   &$T_{cc}$ & 22213.6 & 71082.8 & 52938.8 & 4567.7 & 8491.2 & 7871.3 & 9251.2 \\
   &$p$ & (0.000) & (0.000) & (0.000) & (0.000) & (0.000) & (0.000) & (0.000) \\
   &$b$ & 1.55 & 1.38 & 1.48 & 1.43 & 2.06 & 1.80 & 1.59 \\
   \hline 
 \multirow{4}{*} {Chengdu}
   &$n$ & 9540 & 32121 & 12470 & 869 & 3469 & 4711 & 1606 \\
   &$T_{cc}$ & 30919.1 & 158292.7 & 67452.3 & 4266.7 & 8926.94 &10328.6 & 7816.2 \\
   &$p$ & (0.000) & (0.000) & (0.000) & (0.000) & (0.000) & (0.000) & (0.000) \\
   &$b$ & 1.54 & 1.55 & 1.58 & 1.72 & 1.72 & 1.20 & 1.63 \\
   \hline 
 \multirow{4}{*} {Wuhan}
   &$n$ & 4443 & 18406 & 7684 & 1027 & 3168 & 3103 & 1397 \\
   &$T_{cc}$ & 20633.4 & 99795.5 & 44907.7 & 4176.0 & 10303.7 & 11772.3 & 6427.0 \\
   &$p$ & (0.000) & (0.000) & (0.000) & (0.000) & (0.000) & (0.000) & (0.000) \\
   &$b$ & 1.61 & 1.78 & 1.68 & 1.65 & 1.57 & 1.63 & 1.70 \\
   \hline     
\end{tabular}
\small{\emph{Note: a) Ss. stands for statistics; b)$n$ is the total number of facilities; c)$T_{cc}$ is the statistical score of $\chi^2$; d)$p$ is the p-value of $\chi^2$ test; e) b is the exponent in Taylor's power function.}}
\end{table}

\end{document}